\documentclass{ws-jnopm}
\usepackage{url,overcite}
\usepackage[colorlinks,citecolor=blue,urlcolor=blue]{hyperref}
\usepackage{graphicx}
\usepackage{braket}
\usepackage{amsmath}
\usepackage{subfigure}
\begin{document}

\markboth{Muhammed Ashefas C.H., Manosh T.M. \& Ramesh Babu Thayyullathil}{Preservation of dynamics in coupled cavity system using second order nonlinearity}

%%%%%%%%%%%%%%%%%%% Publisher's Area please ignore %%%%%%%%%%%%%%%%%%%%%%%
\catchline{}{}{}{}{}
%%%%%%%%%%%%%%%%%%%%%%%%%%%%%%%%%%%%%%%%%%%%%%%%%%%%%%%%%%%%%%%%%%%%%%%%%%

\title{Preservation of dynamics in coupled cavity system using second order nonlinearity}

\author{Muhammed Ashefas C.H.$^{*,\dagger,\ddag}$, Manosh T.M.$^{*}$ and Ramesh Babu Thayyullathil$^{*}$}

\address{$^{*}$Department of Physics, Cochin University of Science and Technology, Kochi, Kerala, 682022, India\\$^{\dagger}$Department of Physics, Government Brennen College, Thalassery, Kerala, 670106, India\\
	$^{\ddag}$ashefas@cusat.ac.in}

\maketitle

\begin{history}
\received{(Day Month Year)}
\accepted{(Day Month Year)}
\published{(Day Month Year)}
%\comby{(xxxxxxxxxx)}
\end{history}

\begin{abstract}
	
    We study two coupled cavities with a single two-level system in each and a second-order non-linear process in one of the cavities.  Introduction of a harmonic time dependence on the non-linear coupling is utilized for the preservation of dynamics. It is observed that, even though the preservation period is independent of the initial state, the preserved dynamics depends on the initial state. We calculated the von Neumann entropy and mutual information to study the entanglement present between the subsystems. From which it is found that the time-dependent coupling also preserves the entanglement produced in the system. 

\end{abstract}


\section{\label{sec:level1}Introduction}
Jaynes-Cummings model (JCM) describes the interaction of quantized electromagnetic radiation with a single two-level system (TLS/qubit \cite{PhysRevA.51.2738}) in the rotating wave approximation \cite{1443594}. Over the years JCM has undergone various modifications and has touched different aspects of light-matter physics; including non-linear, deformed and multilevel systems \cite{PhysRevA.45.6816,PhysRevA.50.1785,PhysRev.93.99}. Tailoring this fundamental interaction is the basis of quantum technologies, like, quantum computers\cite{Ladd2010}, quantum computations\cite{doi:10.1142/5528,RevModPhys.68.733} and quantum simulations\cite{RevModPhys.86.153,Blatt2012}. Various optical schemes have been reported in this regard \cite{PhysRevLett.95.010501,PhysRevA.65.042305,PhysRevA.71.060310,Knill2001,Northup2014}. Cavity configurations for generation of entanglement \cite{Angelakis:07,PhysRevA.78.022323,Liu2016,PhysRevB.90.174307,Hacker2019}, preservation of quantum coherence\cite{Man2015,Cheng2016,Mortezapour2018}, recovery of quantum correlations\cite{Orieux2015,Xu2013}, controlled quantum state transfer  \cite{Pfaff2017,Xu2016,Wu2017,doi:10.1142/S0218863518500352},    quantum state squeezing\cite{PhysRevA.50.1867,PhysRevA.37.3175} and ultra strong quantum systems \cite{Stokes2019} are well reviewed in the literature.\\

Nonlinear processes are an integral part of experiments in quantum optics\cite{Caspani2017}. Among which, second order non-linear processes are of special interest in optical squeezing\cite{PhysRevLett.57.2520} and entangled pair production\cite{PhysRevLett.75.4337}. Single photon non-linear process has been experimentally reported using 2D materials like graphene nanostructures \cite{Manzoni_2015} and dichalcogenide MoS$_2$\cite{DinparastiSaleh2018}. This completely transforms the experimental outlook and realization of quantum states.\\

In this article, we study the effects of degenerate second order non-linear process\cite{doi:10.1080/713820226} on the dynamics of quantum states.  Further, the influence of a harmonic time-dependent non-linear coupling is investigated and the emergence of preservation of dynamics with a tunable spontaneous parametric down conversion (SPDC) coupling is shown. The studies were extended to off resonant case.\\ 

Measures such as, entropy, concurrence, fidelity etc. can be used to quantify the quantumness of the system  \cite{PhysRevLett.116.070504,PhysRevA.89.052302,PhysRevA.75.032315,PhysRevLett.113.260502,PhysRevA.79.012318,PhysRevA.94.022324,Luo2017}. Also, measures like continuous variable synchronization \cite{PhysRevLett.111.103605,PhysRevLett.113.154101} and mutual information \cite{PhysRevLett.100.014101,PhysRevA.91.012301,Manzano2013} can estimate the correlations between the systems. In information theory, mutual information is treated as a measure of entanglement \cite{PhysRevE.98.052205}. As the number of subsystems increases, quantifying the correlation between a given pair demands a mutual measure like mutual information. All these measures have been used in various works \cite{PhysRevLett.117.073601,PhysRevLett.121.063601,PhysRevE.89.022913,PhysRevA.91.061401,PhysRevLett.118.243602,PhysRevA.97.013811}. Here, we use von Neumann entropy and mutual information for this purpose.

\section{\label{sec:level2}Coupled cavities with qubits}

We consider two coupled single mode ($\omega_c$) optical cavities with a qubit/two-level system (TLS) of level separation $\omega_{a}$ in both. By taking $\hbar=1$, the system can be described by the sum of free Hamiltonian and the interaction Hamiltonian given by \cite{doi:10.1080/09500340601108851}, 
\begin{equation}
\hat{H}_0=\sum_{i=1}^{2}\left(\frac{1}{2}\omega_{ai}\sigma_z^{(i)}+\omega_{ci}{a_i}^{\dagger}a_i\right),
\label{equation:1}
\end{equation}
and 
\begin{equation}
\hat{H}_I=J\left({a_1}^{\dagger}a_2+a_1{a_2}^{\dagger}\right)+\sum_{i=1}^{2}\lambda_i\left({a_i}^{\dagger}\sigma_-^{(i)}+a_i\sigma_+^{(i)}\right).
\label{equation:2}
\end{equation}

\noindent Here, $\lambda_i$ is the coupling between photons and qubit in the $i$th cavity and $J$ is the photon hopping factor. The operators $a_i$ $(a_i^{\dagger})$ is the annihilation (creation) operator of the $i$th cavity mode, $\sigma_z^{(i)}$ is the population inversion operator, $\sigma_+^{(i)}$ $(\sigma_-^{(i)})$ is the raising (lowering) operator of qubits in the $i$th cavity. A general state, of the form $\ket{\psi}=\ket{\mbox{qubit}_1,\mbox{field}_1,\mbox{qubit}_2,\mbox{field}_2}$,  with a single excitation can be written in the Fock basis as,
\begin{equation}
\ket{\psi(t)}=q_1\ket{1000}+f_1\ket{0100}+q_2\ket{0010}+f_2\ket{0001},
\label{equation:3}
\end{equation}

\noindent where $q_i$ and $f_i$ are the time-dependent coefficients of qubits and fields respectively. The system can be easily solved for resonant case by solving the Schr\"{o}dinger equation,
\begin{equation}
i\frac{\partial\ket{\psi}}{\partial t}=\hat{H}\ket{\psi}.
\label{equation:4}
\end{equation} 
The resulting coupled differential equations are,
\begin{eqnarray}
i\frac{\partial}{\partial t} q_1(t)&&=\lambda f_1(t)\,,
\label{equation:5}\\
i\frac{\partial}{\partial t} f_1(t)&&=\lambda q_1(t)+Jf_2(t)\,,
\label{equation:6}\\
i\frac{\partial}{\partial t} q_2(t)&&=\lambda f_2(t)\,,
\label{equation:7}\\
i\frac{\partial}{\partial t} f_2(t)&&=\lambda q_2(t)+Jf_1(t)\,.
\label{equation:8}
\end{eqnarray}

\noindent Here, we have taken $\lambda_i=\lambda$ and $\omega_{ai}=\omega_{ci}=\omega$. The Eqs. (\ref{equation:5})--(\ref{equation:8}) can be used to obtain the respective Laplace transforms of $q_1(t)$, $q_2(t)$, $f_1(t)$ and $f_2(t)$s as,
\begin{eqnarray}
Q_1(s)=&&\frac{- i J \lambda^{2} q_{2}(0) + J \lambda f_{2}(0) s + i \lambda f_{1}(0) \left(\lambda^{2} + s^{2}\right)}{J^{2} s^{2} + \lambda^{4}+ 2 \lambda^{2} s^{2} + s^{4}}\nonumber\\&&-\frac{q_{1}(0) s \left(J^{2} + \lambda^{2} + s^{2}\right)}{J^{2} s^{2} + \lambda^{4} + 2 \lambda^{2} s^{2} + s^{4}},
\label{equation:9}
\end{eqnarray}
\begin{eqnarray}
F_1(s)=&&\frac{J \lambda q_{2}(0) s + i J f_{2}(0) s^{2} + i \lambda q_{1}(0) \left(\lambda^{2} + s^{2}\right) }{J^{2} s^{2} + \lambda^{4} + 2 \lambda^{2} s^{2} + s^{4}} \nonumber\\&&-\frac{f_{1}(0) s \left(\lambda^{2} + s^{2}\right)}{J^{2} s^{2} + \lambda^{4} + 2 \lambda^{2} s^{2} + s^{4}},
\label{equation:10}
\end{eqnarray}
\begin{eqnarray}
Q_2(s)=&&\frac{- i J \lambda^{2} q_{1}(0) + J \lambda f_{1}(0) s + i \lambda f_{2}(0) \left(\lambda^{2} + s^{2}\right)}{J^{2} s^{2} + \lambda^{4} + 2 \lambda^{2} s^{2} + s^{4}}\nonumber\\&& - \frac{ q_{2}(0) s \left(J^{2} + \lambda^{2} + s^{2}\right)}{J^{2} s^{2} + \lambda^{4} + 2 \lambda^{2} s^{2} + s^{4}},
\label{equation:11}
\end{eqnarray}
\begin{eqnarray}
F_2(s)=&&\frac{J \lambda q_{1}(0) s + i J f_{1}(0) s^{2} + i \lambda q_{2}(0) \left(\lambda^{2} + s^{2}\right) }{J^{2} s^{2} + \lambda^{4} + 2 \lambda^{2} s^{2} + s^{4}}.\nonumber\\&&- \frac{ f_{2}(0) s \left(\lambda^{2} + s^{2}\right)}{J^{2} s^{2} + \lambda^{4} + 2 \lambda^{2} s^{2} + s^{4}}.
\label{equation:12}
\end{eqnarray}

\noindent Now by choosing appropriate initial conditions we can obtain solutions to the Eqs. (\ref{equation:5})--(\ref{equation:8}) by taking the inverse Laplace transform of the Eqs. (\ref{equation:9})--(\ref{equation:12}). For instance, with $\ket{\psi(0)}=\ket{1000}$, we have $q_1(0)=1$ and $f_1(0)=q_2(0)=f_2(0)=0$ and  the corresponding time evolution of probabilities becomes, 
\begin{equation}
|q_{1}(t)|^2=\left| \frac{\left[\left(\xi_+^2+2\lambda^2\right) \cosh \left(\frac{t\xi_-}{\sqrt{2}}\right)-\left(\xi_-^2+2\lambda^2\right) \cosh \left(\frac{t\xi_+}{\sqrt{2}}\right)\right]^2}{4(J^4+4 \lambda ^2 J^2)}\right|,
\label{equation:13}
\end{equation}
\begin{equation}
|q_{2}(t)|^2= \left| \frac{\left[\xi_+ \sinh \left(\frac{t \xi_-}{\sqrt{2}}\right)-\xi_- \sinh \left(\frac{t \xi_+}{\sqrt{2}}\right)\right]^2}{2(J^2+4\lambda^2)}\right|,
\label{equation:14}
\end{equation}
\begin{eqnarray}
|f_{1}(t)|^2=&&\left| \frac{\lambda ^2\left[\sinh ^2\left(\frac{t \xi_-}{\sqrt{2}}\right)+\sinh^2\left(\frac{t \xi_+}{\sqrt{2}}\right)\right]}{J^2+4 \lambda^2}\right.\nonumber\\&&\left.{}+ \frac{-\xi_- \xi_+ \sinh\left(\frac{t \xi_+}{\sqrt{2}}\right) \sinh \left(\frac{t \xi_-}{\sqrt{2}}\right)}{J^2+4 \lambda^2}\right|,
\label{equation:15}
\end{eqnarray}
\begin{equation}
|f_{2}(t)|^2=\left| \frac{\lambda ^2 \left[\cosh \left(\frac{t \xi_-}{\sqrt{2}}\right)-\cosh \left(\frac{t\xi_+}{\sqrt{2}}\right)\right]^2}{J^2+4 \lambda ^2}\right|\,.
\label{equation:16}
\end{equation}

\noindent Where $\xi_\pm=\sqrt{-J^2-2 \lambda ^2\pm\sqrt{J^4+4J^2 \lambda ^2}}$. The evolution of the qubits and fields attributes to the state transfer between the coupled cavities. We have studied similar system in our previous work with Kerr medium as a controller\cite{doi:10.1142/S0218863518500352}. In the following section we introduce a two photon process to the above system.

\section{\label{sec:level3}Coupled cavities with qubits and two photon process}

Two photon processes are widely used for producing entangled photon pairs through spontaneous parametric down conversion (SPDC)\cite{PhysRevLett.75.4337} and to produce squeezed light\cite{PhysRevLett.57.2520}. The experimental realization of single photon SPDC  \cite{Manzoni_2015,DinparastiSaleh2018} makes it possible to have states with a maximum of one excitation in $\chi^{(2)}$ mode and still shows the effect of two photon process. Further, tunable and enhanced nonlinear wave mixing reported\cite{8107820, doi:10.1098/rsta.2016.0313, PhysRevB.92.161406, doi:10.1021/acsnano.5b06110, doi:10.1021/ph500424a} in graphene nanostructures provide platforms for controlled nonlinear processes.

\begin{figure}[h]
	\centering
	\includegraphics[width=6cm]{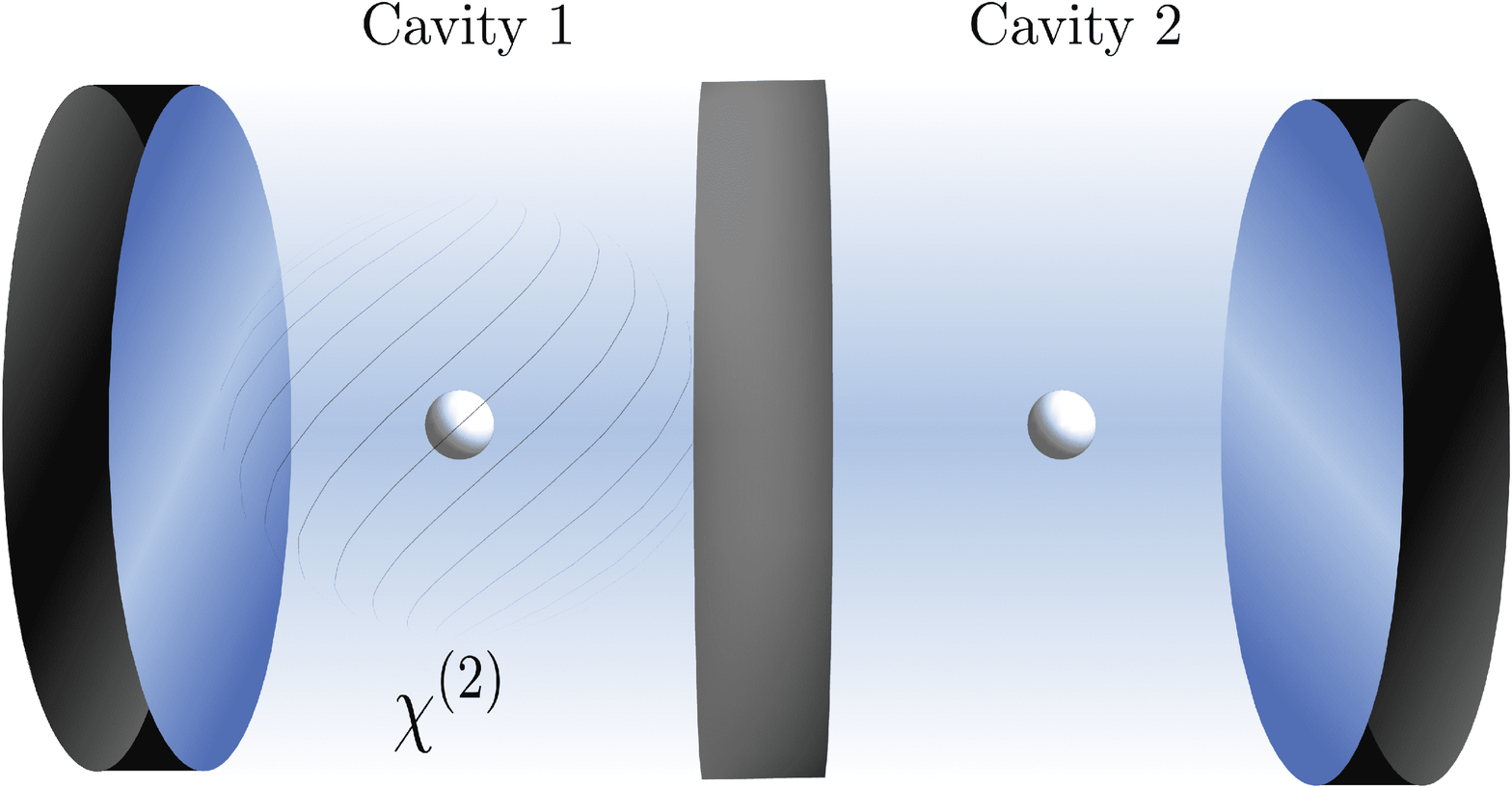}
	\caption{\label{Fig:1}Coupled cavities with a qubit in each and a $\chi^{(2)}$ non-linear medium in the first cavity.}
\end{figure}

Here, a second order nonlinear process is triggered by adding a $\chi^{(2)}$ medium in the first cavity. Where, a single $\omega_b$ photon is converted into two $\omega_c$ photons by degenerate SPDC [see Fig. \ref{Fig:1}]. Now, the Hamiltonian of the system includes an additional term \cite{doi:10.1080/713820226,vedral_2006} $\hat{H}_k$, as, 
\begin{equation}
\hat{H}_k=i\frac{k}{2}\left[{\left({a_1}^{\dagger}\right)}^2b-\left({a_1}\right)^2{b}^{\dagger}\right]+ \omega_{b}b^{\dagger}b.
\label{equation:17}
\end{equation}

\noindent where, $k$ is the non-linear coupling between the $\omega_c$ mode and the $\omega_b$ mode. The operator $b$ $(b^{\dagger})$ is the annihilation (creation) operator for the bosonic mode. By representing $\omega_{b}$ mode as `$\mbox{field}_b$', cavities as `$\mbox{field}_i$' and qubits as `$\mbox{qubit}_i$', an elementary product state may be written as,
\begin{equation}
\ket{\psi}=\ket{\mbox{qubit}_1,\,\mbox{field}_1,\,\mbox{field}_b,\,\mbox{qubit}_2,\,\mbox{field}_2}
\label{equation:18}
\end{equation}

\noindent For a maximum of single excitation in $\omega_{b}$ mode, the general state takes the form,
\begin{eqnarray}
\ket{\psi(t)}=&&\chi(t)\ket{00100}+a(t)\ket{11000}+b(t)\ket{10010}\\\nonumber&&+c(t)\ket{10001}+d(t)\ket{01001}+e(t)\ket{01010}\\\nonumber&&+f(t)\ket{00011}+g(t)\ket{02000}+h(t)\ket{00002}.
\label{equation:19}
\end{eqnarray}

\noindent Now, the dynamics of the system can be studied by solving the following coupled differential equations, which follows from the  Schr\"{o}dinger equation.
\begin{eqnarray}
i\frac{\partial}{\partial t} \chi(t)&&=\omega\chi(t)-i\frac{k}{\sqrt{2}}g(t)\,,\label{equation:20}\\
i\frac{\partial}{\partial t} a(t)&&=\omega a(t) + \sqrt{2}\lambda g(t)+J c(t)\,,\label{equation:21}\\
i\frac{\partial}{\partial t} b(t)&&=\omega b(t) + \lambda e(t)+\lambda c(t)\,,\label{equation:22}\\
i\frac{\partial}{\partial t} c(t)&&=\omega c(t) + \lambda d(t)+\lambda b(t)+Ja(t)\,,\label{equation:23}\\
i\frac{\partial}{\partial t} d(t)&&=\omega d(t) + \lambda c(t)+\lambda e(t)+\sqrt{2}Jh(t)+\sqrt{2}Jg(t)\,,\label{equation:24}\\
i\frac{\partial}{\partial t} e(t)&&=\omega e(t) + \lambda d(t)+\lambda b(t)+Jf(t)\,,\label{equation:25}\\
i\frac{\partial}{\partial t} f(t)&&=\omega f(t) + \sqrt{2}\lambda h(t)+Je(t)\,,\label{equation:26}\\
i\frac{\partial}{\partial t} g(t)&&=\omega g(t) + \sqrt{2}\lambda a(t)+\sqrt{2}Jd(t)+i\frac{k}{\sqrt{2}}\chi(t)\,,\label{equation:27}\\
i\frac{\partial}{\partial t} h(t)&&=\omega h(t) + \sqrt{2}\lambda f(t)+\sqrt{2}Jd(t)\,.\label{equation:28}
\end{eqnarray}
In the above coupled equations we have taken, $\omega_{b}=2\omega_{c}=2\omega_{a}=2\omega$ and $\lambda_i=\lambda$. The Eqs. (\ref{equation:20})--(\ref{equation:28}) can be solved using the Laplace transform method used earlier. Here, for a given set of initial conditions, we solve the differential equations numerically \cite{JOHANSSON20121760}. 

\subsection*{Case 1. Single excitation in $\omega_{b}$}

We start with a single excitation of $\omega_{b}$ mode, $\ket{\psi(0)}=\ket{00100}$, which undergoes degenerate SPDC resulting in two $\omega_c$ photons. These photons can interact with the qubits through the coupling $\lambda$, tunnel between the cavities, for non zero values of $J$ and also undergo up conversion to $\omega_{b}$. The population inversion $\braket{\sigma_z}$, for qubits ($Q1$ and $Q2$), harmonically oscillates between $-1$ (ground state $\ket{0}$) and zero (superposed state $\left(\ket{0}+\ket{1}\right)/\sqrt{2}$). Behaviour of $\sigma_z$ suggest a correlation between the qubits [see Fig. \ref{Fig:2a}]. Here, the overall harmonic behaviour of the qubits does not change as the time passes by. \\

\subsection*{Case 2. Entangled initial state, $\left(\ket{00100}+\ket{01001}\right)/\sqrt{2}$}

\begin{figure*}[h]
	\centering
	\subfigure[]{\label{Fig:2a}\includegraphics[width=0.49\textwidth]{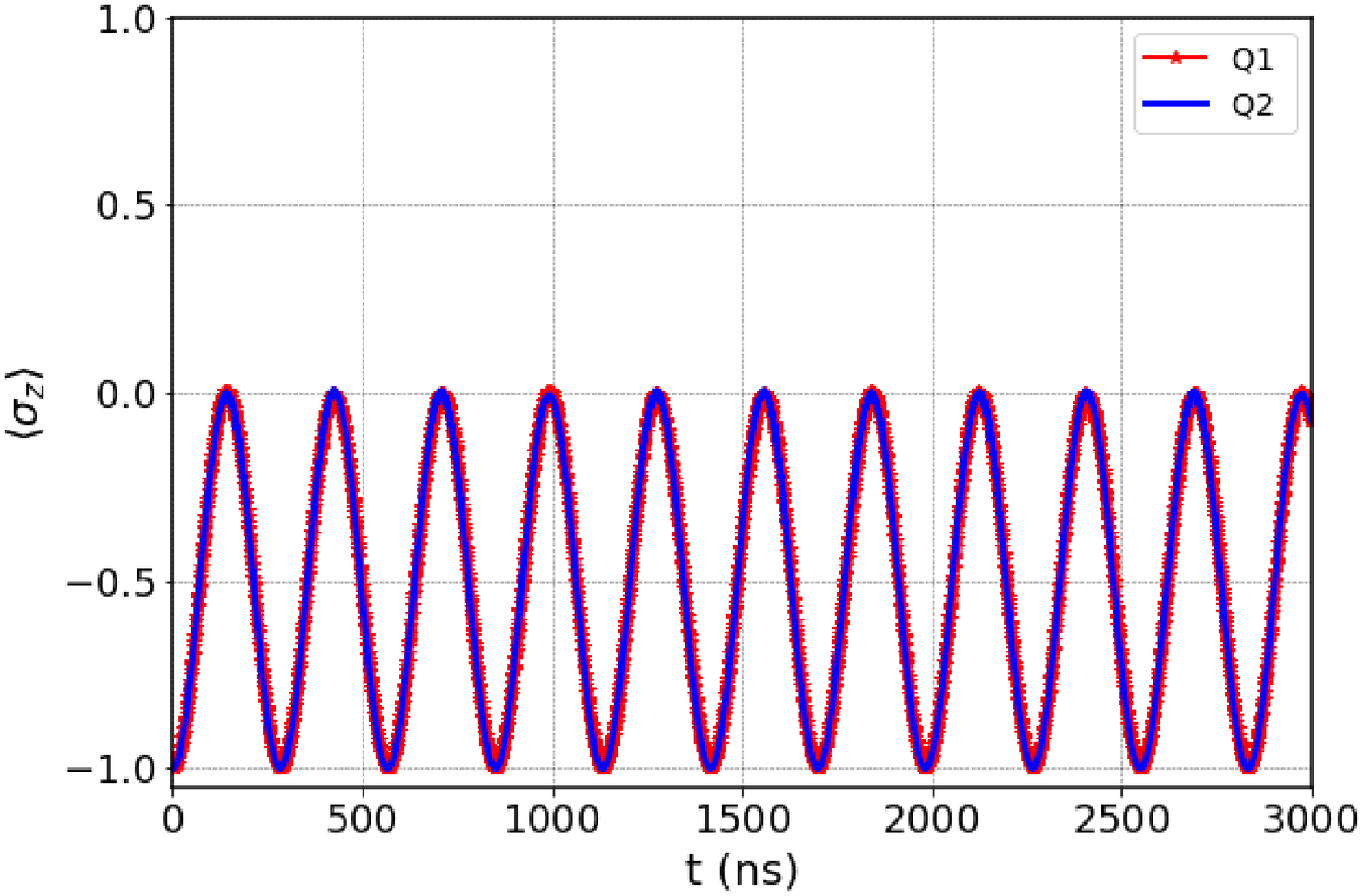}}
	\subfigure[]{\label{Fig:2b}\includegraphics[width=0.49\textwidth]{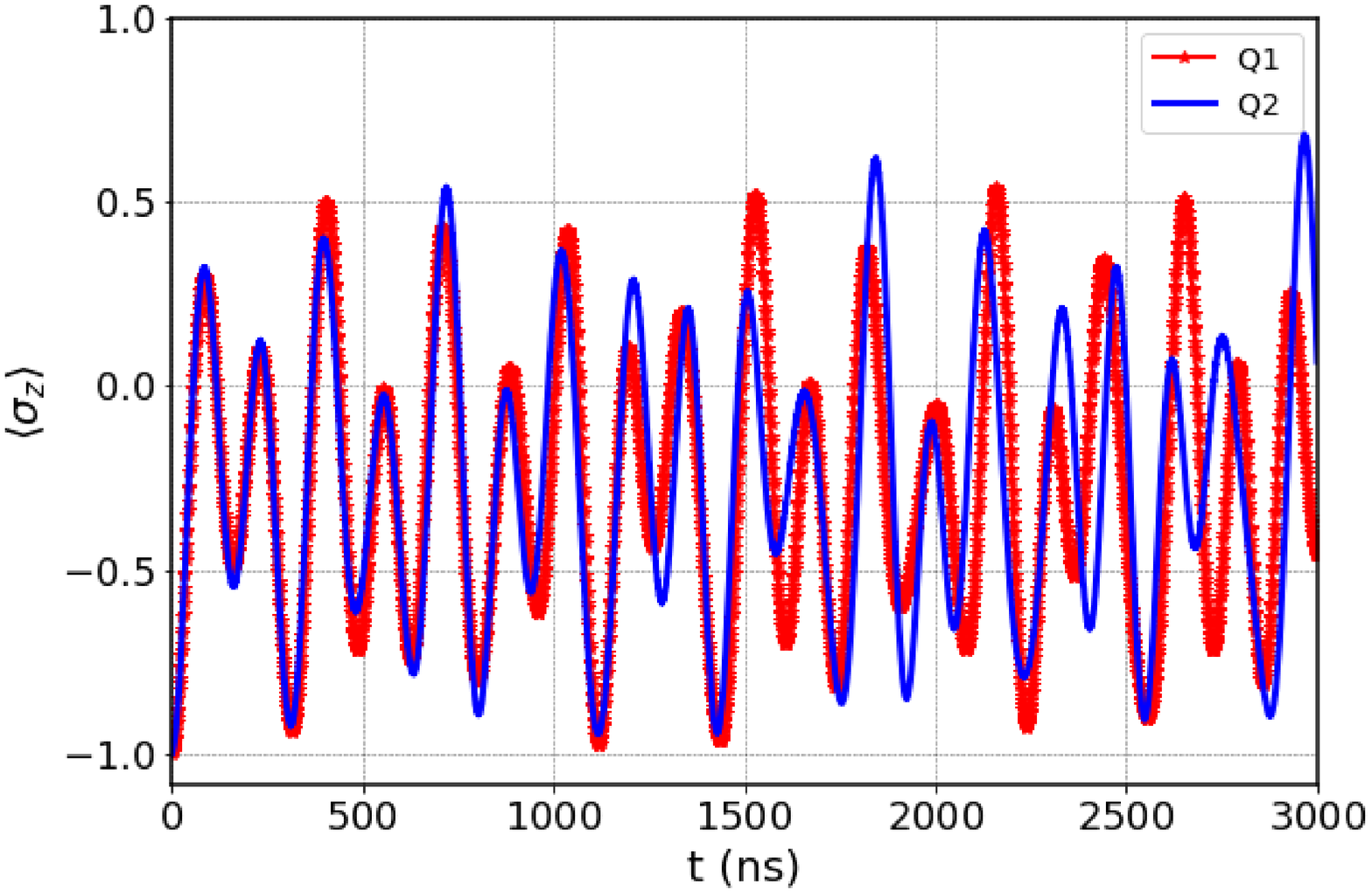}}
	\caption{\label{Fig:2}Population inversion, $\braket{\sigma_z}$, of qubit 1 (red colour) and qubit 2 (blue colour) for $\omega=10\times2\pi$ GHz, $\lambda=0.1\,\omega$, $J=0.05\,\lambda$, $k=0.010\,\omega$ and (a) $\ket{\psi(0)}=\ket{00100}$ (b) $\ket{\psi(0)}=\left(\ket{00100}+\ket{01001}\right)/\sqrt{2}$ .}
\end{figure*}

To illustrate the dependency of nonlinear coupling, we repeat it for a different initial state and $k$ values. For $\ket{\psi(0)}=\left(\ket{00100}+\ket{01001}\right)/\sqrt{2}$, where the subsystems are entangled, the system does not preserve the identical behaviour as exhibited in the previous case. This means the characteristics of temporal evolution of qubits depends more on the initial state rather than it depends on the coupling factors. For the numerical analysis we have taken $\omega=10\times2\pi$ GHz, $\lambda=0.1\,\omega$ and $J=0.05\,\lambda$ [see Fig. \ref{Fig:2b}].

\begin{figure*}[h]
	\centering
	\includegraphics[width=1\textwidth]{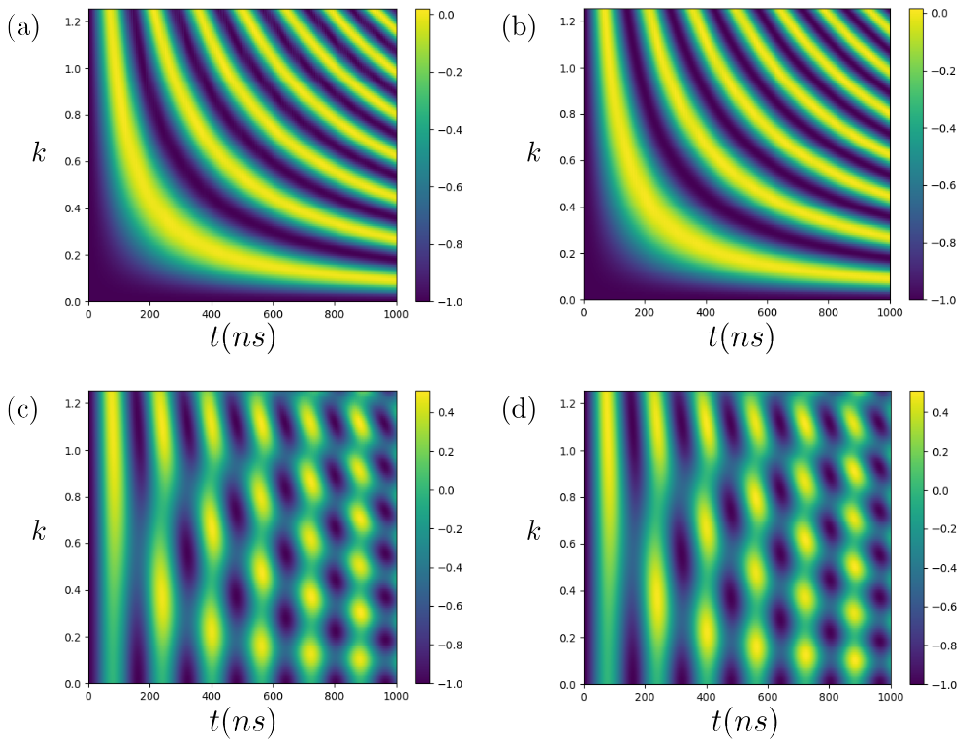}
	\caption{ \label{Fig:3}Population inversion for different $k$ values (a) qubit 1 and (b) qubit2 for the initial state $\ket{00100}$. (c) qubit 1 and (d) qubit 2 for the initial state $\left(\ket{00100}+\ket{01001}\right)/\sqrt{2}$, $\Delta = 0$.}
\end{figure*}

\begin{figure*}[h]
	\centering
	\includegraphics[width=1\textwidth]{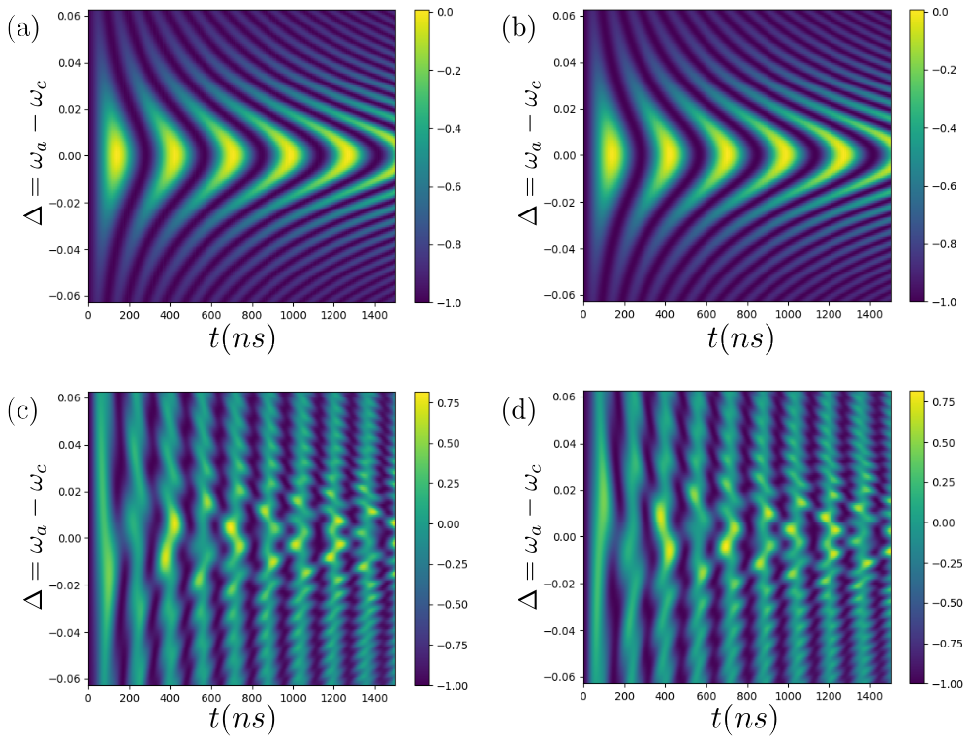}
	\caption{ \label{Fig:4}Population inversion for different detuning $\Delta$ (a) qubit 1 and (b) qubit2 for the initial state $\ket{00100}$. (c) qubit 1 and (d) qubit 2 for the initial state $\left(\ket{00100}+\ket{01001}\right)/\sqrt{2}$, $k = 0.010\omega$.}
\end{figure*}

Changing the value of $k$ affects the oscillation irrespective of the initial states. Further we spanned $k$ over a range and studied the dynamics. The results are shown in Fig. \ref{Fig:3}. From the figures, it is evident that various values of coupling factors can bring changes in the dynamics of the system. For the first case, an increase in $k$ increases the oscillation rate. But the amplitude of the system remains unaffected. On the other hand, in the second case, we can see responses in the amplitude of the oscillations and the dynamics does not always carries a simple harmonic behaviour. In the second case, the qubits behaves rather differently, as the time passes by. \\

The response of the system will depend on the detuning present in the configuration. To confirm this, we extend our studies to off resonant cases. Till now, the cavity mode ($\omega_{c}$) and the atomic transition frequency ($\omega_{a}$) were same. Thus the detuning, $\Delta = \omega_{a}-\omega_{c}$ was zero. Figure \ref{Fig:4} shows the sensitivity of the system to detuning. As detuning increases, the amplitude of population inversion is reduced for both cases. Similar to the variation of $k$ in the first case, detuning affects the oscillation rate. Despite the nonlinear process being confined to one cavity, for the special case with $\ket{\psi(0)}=\ket{00100}$, both positive and negative $\Delta$ yields identical results. But the asymmetry in the configuration can be seen in the second case, where we can clearly distinguish between positive and negative detuning as well as qubit 1 and qubit 2. To elaborate further on the temporal characteristics of qubits, we will investigate other measure such as entropy and mutual information in the later section. \\

In conclusion, various values for coupling constants produce notable differences in the dynamics of the system. Hence, a tunable configuration brings the freedom to tailor the dynamics according to the need. Variation in coupling factor can be introduced by making it time dependent. Time dependent coupling schemes has been addressed in cavity systems before \cite{PhysRevA.48.2276}. In the next section we introduce a tunable time dependent nonlinear coupling and investigate the outcome of such a phenomenological model.

\section{\label{sec:level4}Time dependent nonlinear coupling}

Tunable and enhanced non-linearities have been reported in various composite\cite{doi:10.1021/nl300084j, Lin2017, 8107820, doi:10.1098/rsta.2016.0313, PhysRevB.92.161406, doi:10.1021/acsnano.5b06110, doi:10.1021/ph500424a}, especially 2D nano materials. This allows us to theoretically propose a tunable harmonic time dependence on the nonlinear coupling factor $k$. For this purpose, we propose a harmonic time dependence as, 
\begin{equation}
k(t)\equiv k_0\left[1+\sin\left(\Omega t\right)\right],
\label{equation:29}
\end{equation} 

\noindent Here $\Omega$ can vary from zero to infinity, making it as the tunable parameter. When $\Omega=0$ we get the constant coupling as earlier. Table (\ref{tab:1}), gives the numerical values connecting $t$, $\Omega$ and $k$ in the first cycle. We look at the signature of the time dependence of $k$ over the interval in which the value of $k$ decreases from $10\%$ of the maximum value of $k$ to zero and further increases to $10\%$ of the maximum value of $k$. Previously, as $k$ is increased, we have seen an increase in the oscillatory behaviour of qubits for case 1 and non periodic variation in the inversion amplitude for case 2. By tuning the nonlinear coupling we could surpass this and achieve much stable dynamics for a desired period of time. The choice of when to reduce or increase $k$ depends on $\Omega$. And this gives the freedom to tune and control the system with other coupling factor ($\lambda$ and $J$). From Table (\ref{tab:1}), as $\Omega$ increases, there is a shift as well as a reduction in the period over which, the value of $k$ decreases from $10\%$ of the maximum value of $k$ to zero and further increases to $10\%$ of the maximum value of $k$. In principle we could independently modulate these two features by adding other harmonic terms. \\

\begin{table}[h]
\tbl{Numerical values connecting $t$, $\Omega$ and $k$}
{\begin{tabular}{lrrr}
\toprule
$\Omega$&Time at $10\%k_0$&Time at $k=0$&Time at $10\%k_0$\\
(GHz)&(ns)&(ns)&(ns)\\
\colrule
0.000000&-- & --&--\\
0.002222& 1917.7& 2120.6&2323.5\\
0.004444& 958.8& 1060.3&1161.8\\
0.006667& 639.1& 706.8&774.5\\
0.008889& 479.3& 530.1&580.9\\
\botrule
\end{tabular}}\label{tab:1}
\end{table}

\begin{figure}[h]
	\centering
	\subfigure[]{\label{Fig:5a}\includegraphics[width=0.48\textwidth]{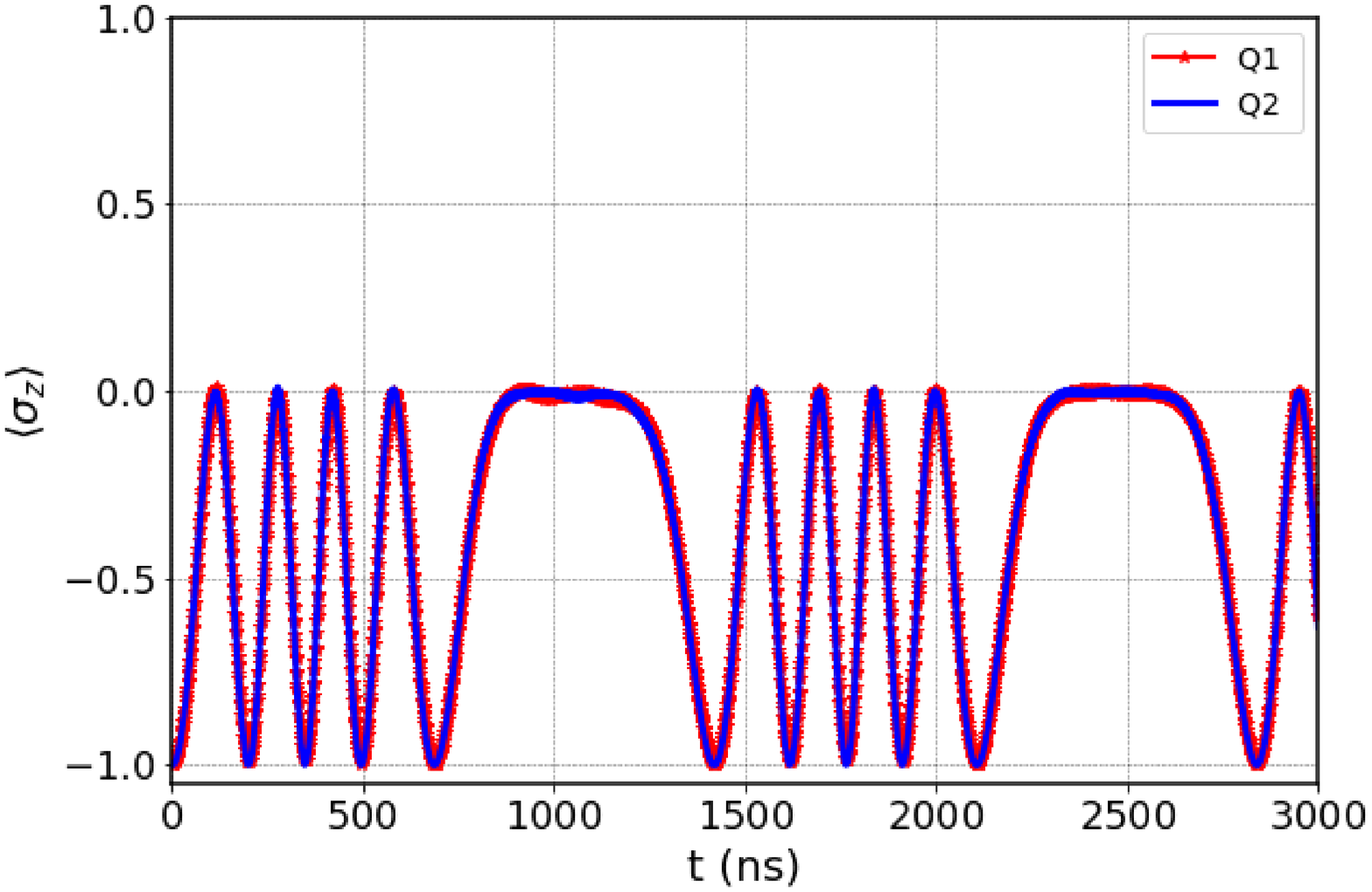}}
	\subfigure[]{\label{Fig:5b}\includegraphics[width=0.48\textwidth]{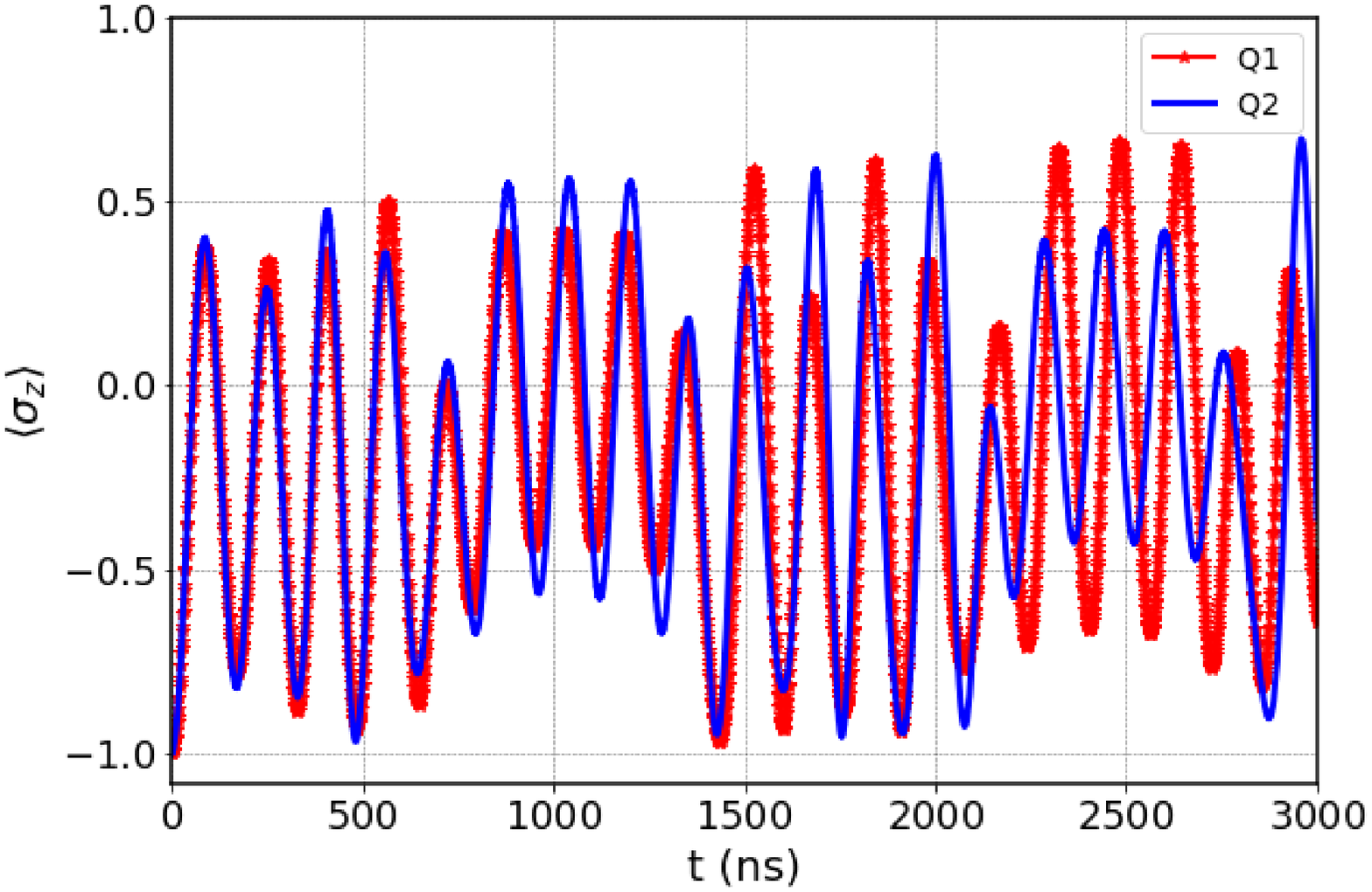}}
	\caption{\label{Fig:5}Population inversion $\braket{\sigma_z}$, of qubit 1 (red colour) and qubit 2 (blue colour) for $\omega=10\times 2\pi$ GHz, $\lambda=0.1\omega$, $J=0.05\lambda$, $\Omega=0.004444$ GHz and $k_0=0.01\omega$. (a)  $\ket{\psi(0)}=\ket{00100}$ (b)  $\ket{\psi(0)}=\frac{1}{\sqrt{2}}\left(\ket{00100}+\ket{01001}\right)$.}
\end{figure}

\noindent The population inversion of the qubits, for the initial state $\ket{00100}$ and $\left(\ket{00100}+\ket{01001}\right)/\sqrt{2}$ and $\Omega=0.004444$ GHz are shown in the Figs. \ref{Fig:5a} and \ref{Fig:5b}.

\begin{figure}
	\centering
	\includegraphics[width=\textwidth]{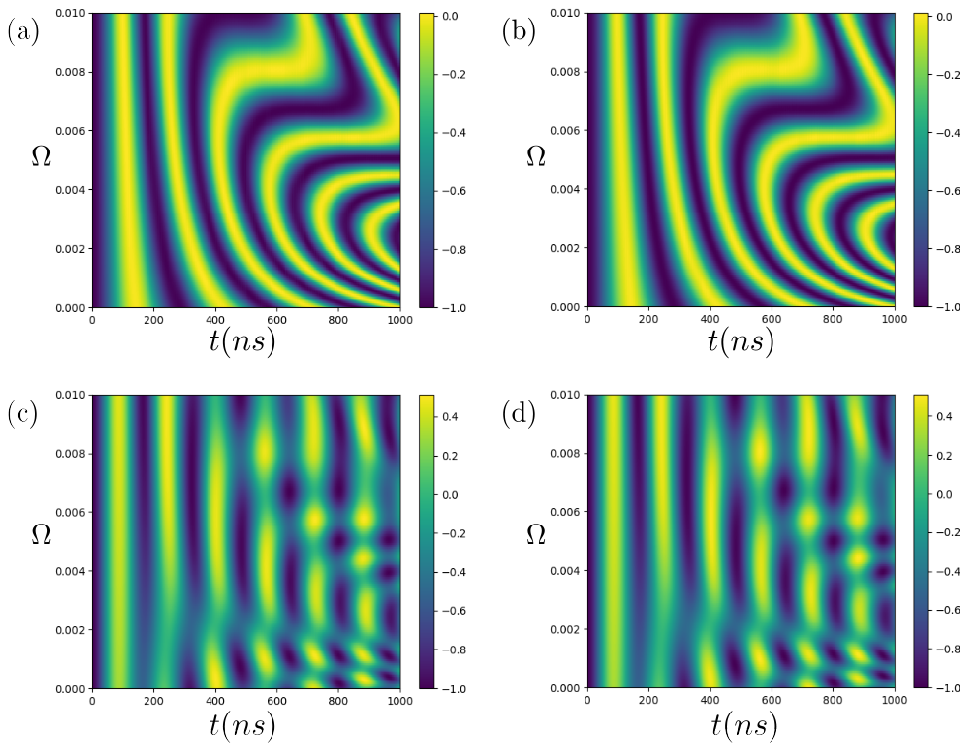}
	\caption{\label{Fig:6} Effect of various time dependent $k$ on population inversion of qubit in cavities. (a) and (b) qubit in the cavities 1 and 2 for the initial state $\ket{00100}$. (c) and (d) qubit in the cavities 1 and 2 for the initial state $\frac{1}{\sqrt{2}}\left(\ket{00100}+\ket{01001}\right)$, $k_0=0.010\omega_c$, $\Delta = 0$.}
\end{figure}

\noindent The figures clearly shows a preservation plateaus around the period 950 $\sim$ 1200 $ns$ for both cases [see Table. (\ref{tab:1})]. Hence, we can say that this phenomena is independent of the initial state. It is interesting to note that, we could preserve different dynamics [see Fig. \ref{Fig:5b}] during the next similar interval. Thus the preservation is also independent of the state at which the preservation period starts. To better understand this, we repeated the calculation for $\Omega$ ranges from $0$ to $0.01$ GHz. Figure \ref{Fig:6} shows the dynamics of qubits 1 and 2 for a range of $\Omega$ and  different initial states with $k_0=0.01\omega$. With $\ket{\psi(0)}=\ket{00100}$, when $\Omega=0$, we see the inversion as it was without the time dependent coupling. When $\Omega$ increases, value of $k$ also increases. In the previous section, for the same initial conditions, an increase in $k$ resulted in increased oscillatory rate. This is clearly seen in time dependent case as well.

After reaching the maximum, value of $k$ will start to descend. This must result in the reduction of oscillation rate. In the Fig. \ref{Fig:6}, we can identify this reduction by the increase in the width of the colour bands.  The dynamics gets more interesting when the value of $k$ decreases from $10\%$ of the maximum value of $k$ to zero and further increases to $10\%$ of the maximum value of $k$. We can see preservation plateaus appearing in the figure. The analysis clearly shows that this plateaus appears for the all ranges of $\braket{\sigma_z}$. For example, in the Fig. \ref{Fig:6} (a), when $\Omega\approx0.0067$ GHz, a preservation plateau appears for the time interval approximately around 650 ns to 750 ns, with $\braket{\sigma_z}\approx-1$. Similarly, when $\Omega\approx0.006$ GHz, a preservation plateau appears for the time interval approximately around 750 ns to 850 ns, with $\braket{\sigma_z}\approx-0.2$. After this preservation period, when $k$ starts to increase, the dynamical behaviour revive and goes on till the next preservation period arrives. These preservation slots can also be seen for $\ket{\psi(0)}=\left(\ket{00100}+\ket{01001}\right)/\sqrt{2}$ ,where, a harmonic oscillatory behaviour is preserved. Such that the population inversion oscillates between, a particular range harmonically [see Figs. \ref{Fig:5b}, \ref{Fig:6}(c) and \ref{Fig:6}(d)].   
 
\begin{figure}
	\centering
	\includegraphics[width=\textwidth]{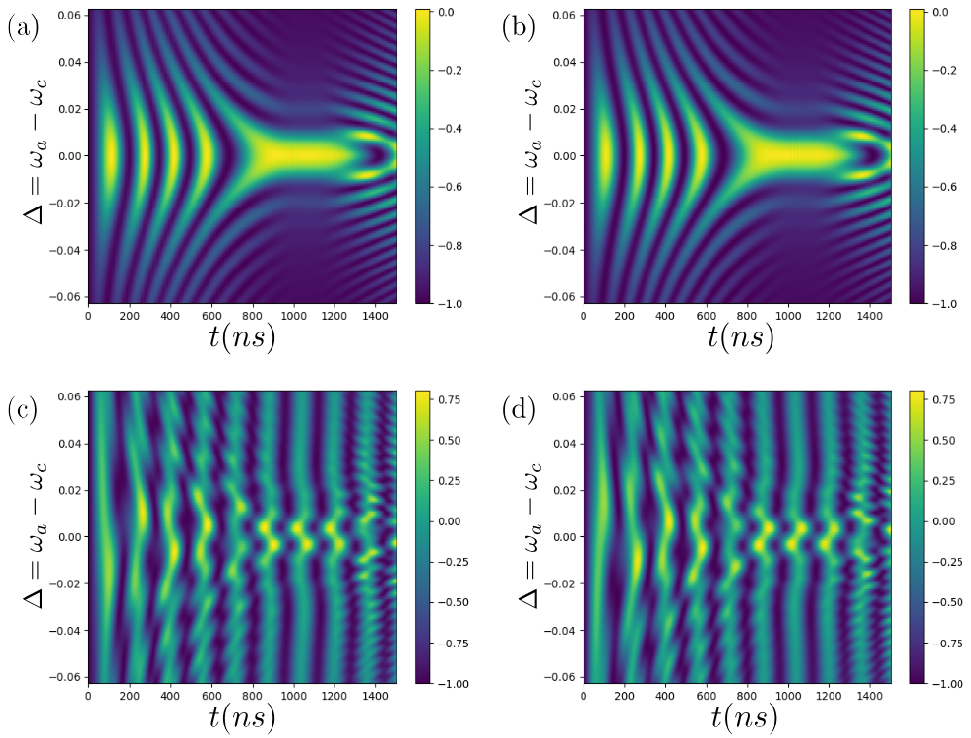}
	\caption{\label{Fig:7} Effect of detuning in time dependent $k$ on population inversion of qubit in cavities. (a) and (b) qubit in the cavities 1 and 2 for the initial state $\ket{00100}$. (c) and (d) qubit in the cavities 1 and 2 for the initial state $\frac{1}{\sqrt{2}}\left(\ket{00100}+\ket{01001}\right)$, $\Omega=0.004444$ GHz and $k_0=0.010\omega_c$.}
\end{figure}  

The system is further studied for off resonant region, with $\omega_{a}\neq\omega_{c}$. Even in the time dependent coupling scheme, detuning affects the dynamics. Figure \ref{Fig:7} shows the results on the new configuration. The figure also highlights the preservation slots for both initial states. As in the time independent case, here we can see identical behaviour for both positive and negative detuning for the special case with initial state as $\ket{00100}$. For the next case, this symmetry is again not seen.\\

This indicates that, whenever we need to control a certain behaviour, we can do so by controlling the coupling factors. In this particular case, we have seen that, a time dependent nonlinear coupling preserves the dynamics of qubits for a tunable period. To address the characteristics of qubits, we further analyse the system. Using mutual information and von Neumann entropy to quantify the quantumness of system. In the next section, we present an overlook of these measures and calculate the same for our system. 

\section{\label{sec:level5} Mutual Information and von Neumann entropy}

A probabilistic theory cannot yield deterministic results due to the lack of information. Hence, one has to account for this lack of information and deduce the physics. Works by Shannon \cite{doi:10.1002/j.1538-7305.1948.tb01338.x} and Neumann \cite{1955mfqm.book.....V} clearly formulate this lack of information as the entropy of the system. It can account, not only the classical probabilities but also the quantumness present in the configuration. Here we use mutual information (mutual entropy) and von Neumann entropy. 

Mutual information (MI) quantifies the amount of information shared between two systems \cite{PhysRevA.71.063821}, and is defined as:
\begin{equation}
I(X:Y)=H(X)+H(Y)-H(X,Y)\,.
\label{equation:30}
\end{equation} 
\noindent Where $H(X)$ and $H(Y)$ is the Shannon entropy and $H(X,Y)$ is the joint Shannon entropy. Classically, the above equation corresponds to the relative entropy with,
\begin{equation}
H(X,Y)\geq H(X)\,\mbox{or}\, H(Y).
\label{equation:31}
\end{equation}
\noindent The quantum analogue of Shannon entropy is the von Neumann entropy \cite{RevModPhys.81.865} ($S$), from which the quantum analogue for mutual information takes the form, 
\begin{equation}
I(m:n)=S\left(\rho_{m}\right)+S\left(\rho_{n}\right)-S\left(\rho_{mn}\right),
\label{equation:32}
\end{equation}
\noindent where $\rho$ is the density matrix and $m, n$ are the indices of the subsystems (sub Hilbert-spaces). Here, $\rho_{mn}$ and $\rho_{m}$ corresponds to the composite system and subsystem respectively. One can find $\rho_{m}$ by taking the partial trace of $\rho_{mn}$ over $n$. Now the inequality becomes,
\begin{equation}
S\left(\rho_{mn}\right)\geq S\left(\rho_{n}\right)
\label{equation:33}
\end{equation} 
\noindent The violation of the above inequality [Eq. (\ref{equation:33})] indicates the presence of entanglement in the system. A general expression for entropy in quantum mechanics is given as,
\begin{equation}
S_{\alpha}\left(\rho_m\right)=\left(1-\alpha\right)^{-1}\log\mbox{Tr}\left(\rho_m\right)^{\alpha},
\label{equation:34}
\end{equation} 
\noindent This is known as the $\alpha$-entropy or the R\'{e}nyi $\alpha$-entropy. Equation (\ref{equation:34}) reduces to von Neumann entropy for the limit $\alpha\to1$ as, 
\begin{equation}
S\left(\rho_m\right)=-\mbox{Tr}\left(\rho_m\log_2\rho_m\right).
\label{equation:35}
\end{equation} 
\noindent Expressions in Eqs. (\ref{equation:32}) and (\ref{equation:35}) are used for our analysis. We numerically compute $\rho$ as a function of time and then calculate the entropies to better understand the quantum states of the system.

\subsection{\label{sec:level5sub1}Von Neumann Entropy}

For von Neumann entropy, first we numerically compute $\ket{\psi(t)}$ using the initial conditions and the time dependent Hamiltonian with the time dependent coupling in Eq. (\ref{equation:29}), and thus obtain $\rho(t)$. Since we are interested in the dynamics of qubits, we trace over the rest of the subsystems and found $\rho_{q1}$ and $\rho_{q2}$. Finally, using the Eq. (\ref{equation:32}), we computed the von Neumann entropy of the qubits, with initial conditions  $\ket{\psi(0)}=\ket{00100}$ and  $\ket{\psi(0)}=\left(\ket{00100}+\ket{01001}\right)/\sqrt{2}$.

\begin{figure}[ht]
	\centering
	\subfigure[]{\label{Fig:8a}\includegraphics[width=\textwidth]{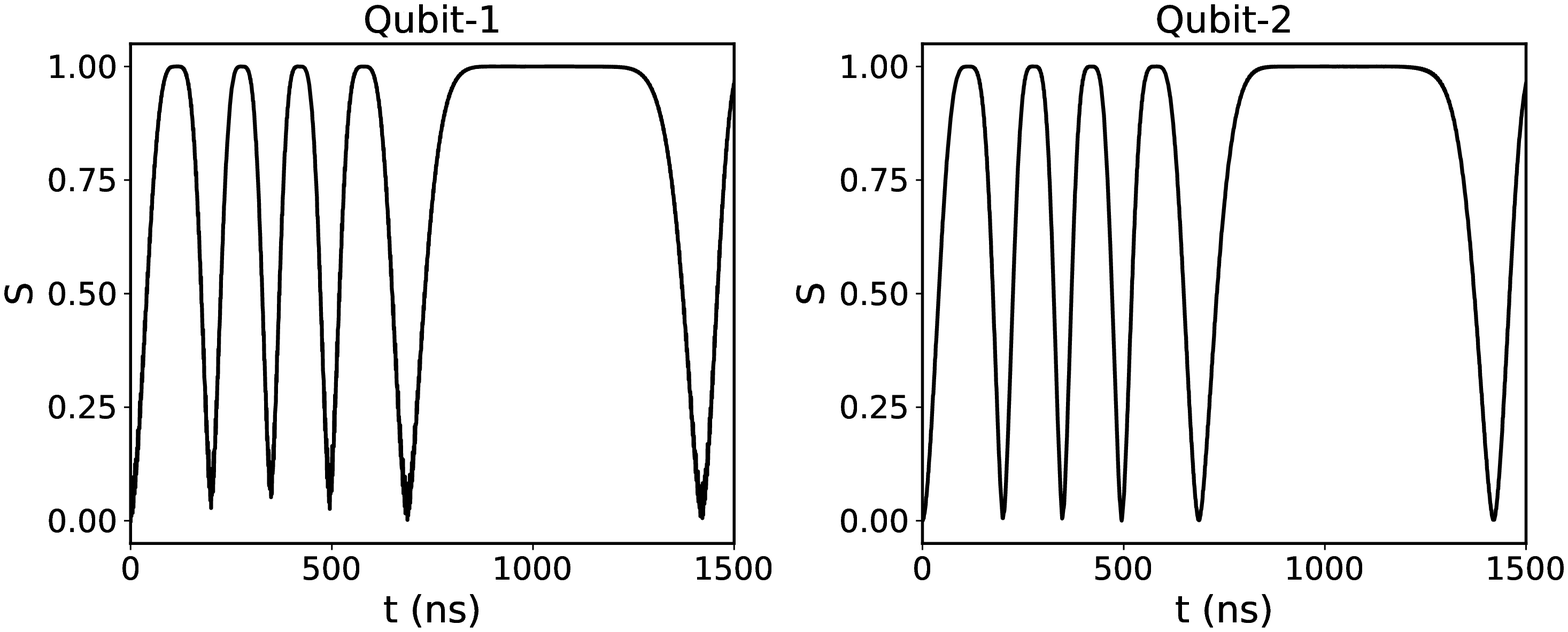}}
	\subfigure[]{\label{Fig:8b}\includegraphics[width=\textwidth]{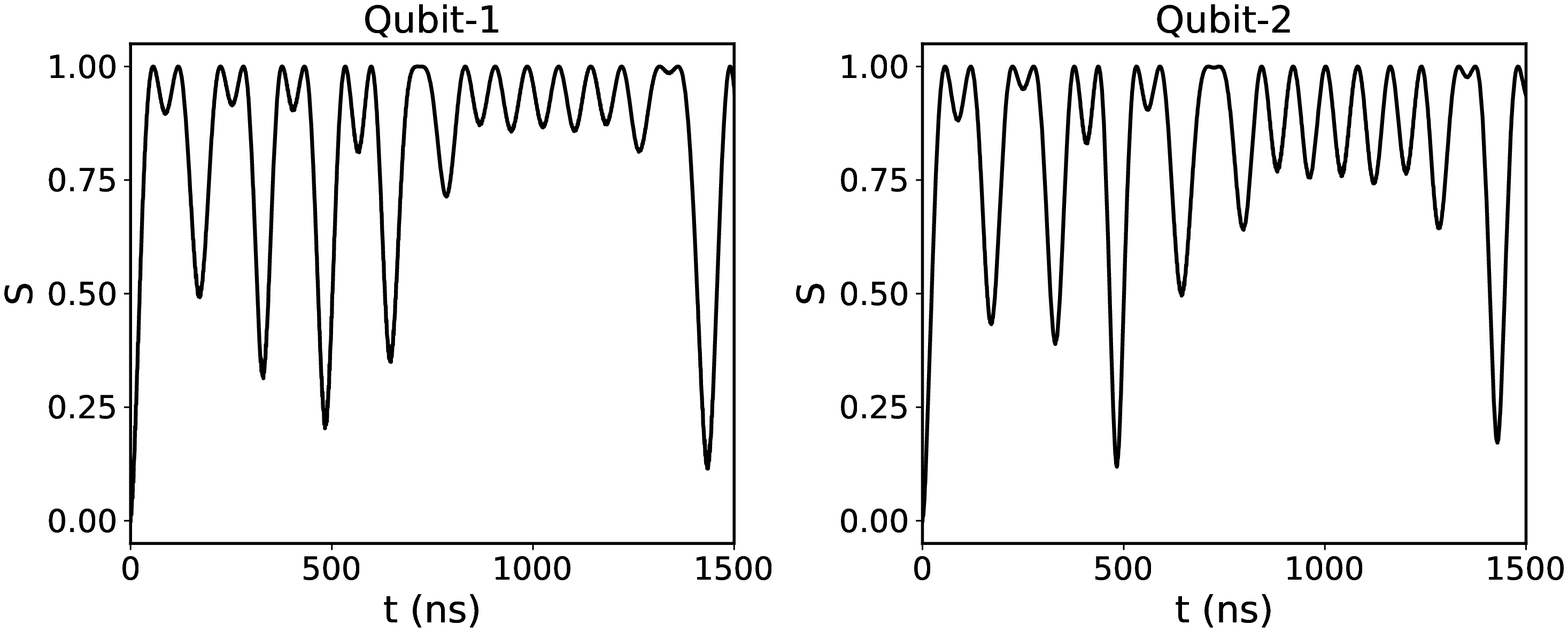}}
	\caption{\label{Fig:8}Von Neumann entropy (a)  $\ket{\psi(0)}=\ket{00100}$ and (b)  $\ket{\psi(0)}=\frac{1}{\sqrt{2}}\left(\ket{00100}+\ket{01001}\right)$. }
\end{figure}

\noindent The results of both initial states are shown in Figs. \ref{Fig:8a} and \ref{Fig:8b}. For the first case, with $\ket{\psi(0)}=\ket{00100}$, the qubits reaches the maximum possible value of $S$ and tends to remain in that state during the preservation period. Since the states are not mixed, this maximum value of von Neumann entropy corresponds to a maximally entangled state ($S=1$). For a different value of $\Omega$, one may preserve a non maximal entanglement ($1>S>0$) or a pure state ($S=0$).

\begin{figure}[ht]
	\centering
	\subfigure[]{\label{Fig:9a}\includegraphics[width=\textwidth]{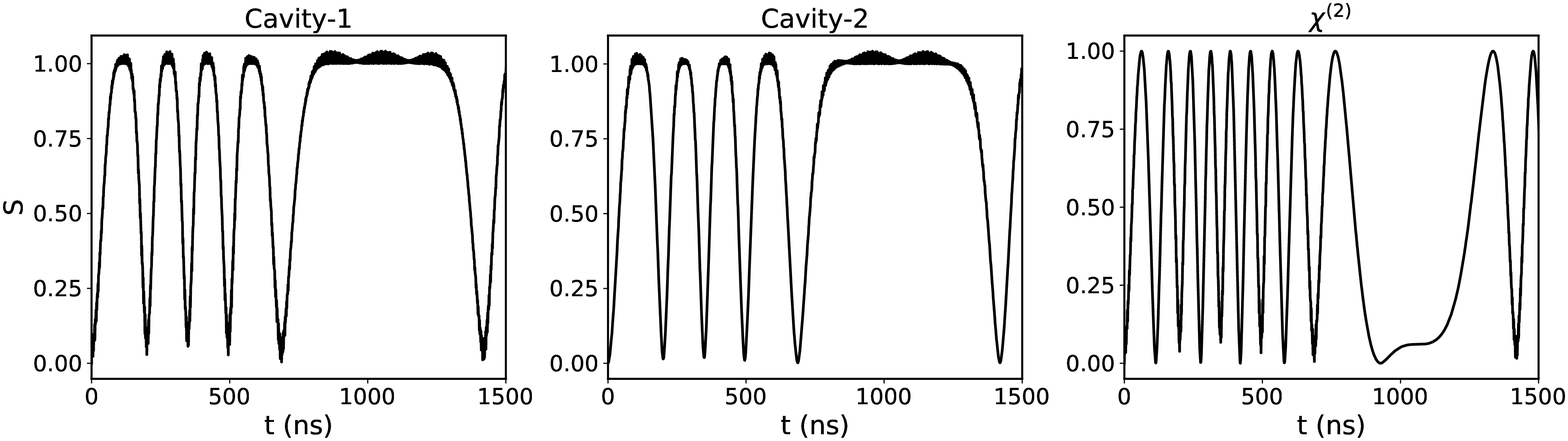}}
	\subfigure[]{\label{Fig:9b}\includegraphics[width=\textwidth]{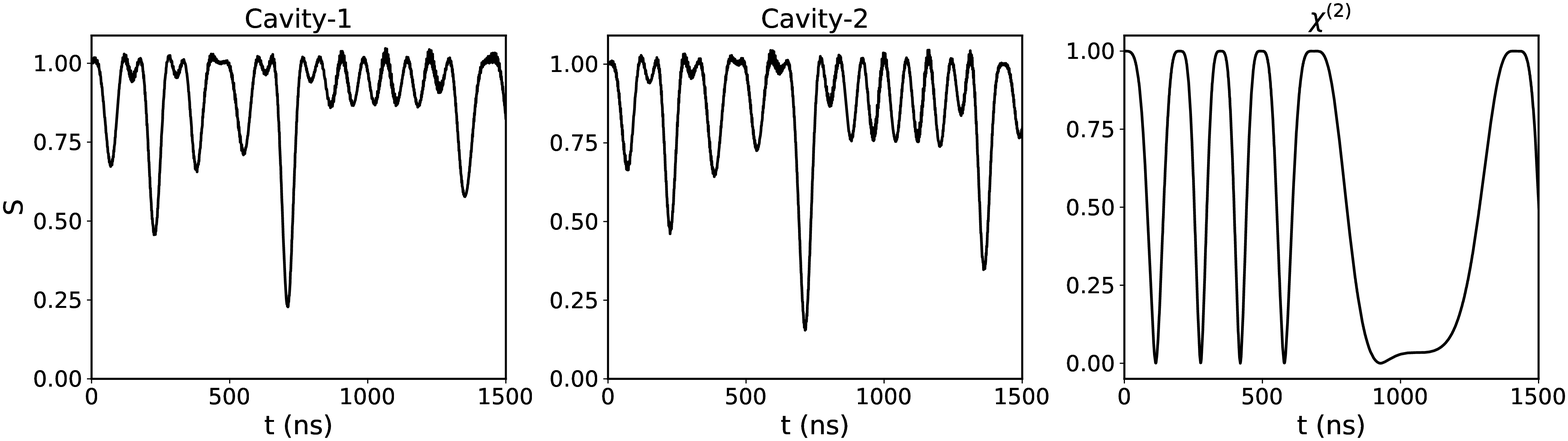}}
	\caption{\label{Fig:9}Von Neumann entropy (a)  $\ket{\psi(0)}=\ket{00100}$ and (b)  $\ket{\psi(0)}=\frac{1}{\sqrt{2}}\left(\ket{00100}+\ket{01001}\right)$. }
\end{figure}

For the second case with $\ket{\psi(0)}=\left(\ket{00100}+\ket{01001}\right)/\sqrt{2}$, the qubits tends to preserve the dynamics during the preservation slot. Before the preservation slot, the von Neumann entropy of the qubits oscillates with larger amplitude difference. But during the time, the oscillation amplitude remains rather consistent. This confirms the preservation of the dynamics of the qubits. Further we did the same analysis on other subsystems present in the configuration and the results are show in Fig.\ref{Fig:9}.

\noindent Here, the small fluctuations in the preservation plateau is due to other coupling factors, such as photon hopping in the system. It has to be note that, the von Neumann entropy of the cavity modes and $\chi^{(2)}$ mode started from a non zero value in the second case [see Fig. \ref{Fig:9b}]. This clearly indicates the initial entanglement between the subsystems. But they are not the maximum, this is because of the sharing of entanglement between the cavity modes. Even though we could infer the entanglement of a subsystem, von Neumann entropy cannot reveal about which subsystem it is entangled to.  To address this one must calculate a measure, which incorporate two subsystems at once. Here we use mutual information, defined in Eq. (\ref{equation:32}), as our measure for the same.

\subsection{\label{sec:level5sub2}Mutual Information}

Earlier, we found that there is an entanglement present between our subsystems. In this section, we compute the mutual information between each pairs of subsystems to understand the distribution of information. Mutual information is measured between two subsystem, since we have five subsystems, we need to compute the partially reduced density matrices, such that each pairs are accounted. For example, to study mutual information between qubit 1 and qubit 2, we trace out the cavities and other bosonic modes from the complete density matrix, ($\rho$) such that we end up with a partially reduced density matrix ($\rho_{q1q2}$). We also find the completely reduced density matrix $\rho_{q1}$ and $\rho_{q2}$. Now we calculate the von Neumann entropies and compute mutual information as, 
\begin{equation}
I(q1:q2)=S\left(\rho_{q1}\right)+S\left(\rho_{q2}\right)-S\left(\rho_{q1q2}\right),
\end{equation}

For this, we choose our initial state and compute the density matrices as a function of time. Further, the reduced ($\rho_{m}$) and partially reduced ($\rho_{mn}$) density matrices were computed, after which we proceed to compute mutual information as discussed above. The procedure is repeated for all such pairs in the configuration. Here, the initial states considered are $\ket{\psi(0)}=\ket{00100}$ and $\ket{\psi(0)}=\left(\ket{00100}+\ket{01001}\right)/\sqrt{2}$ with $\Omega=0.004444$ GHz. For consistency, the coupling constants retains the previously assigned values. 

\begin{figure}[h]
	\centering
	\subfigure[]{\label{Fig:10a}\includegraphics[width=\textwidth]{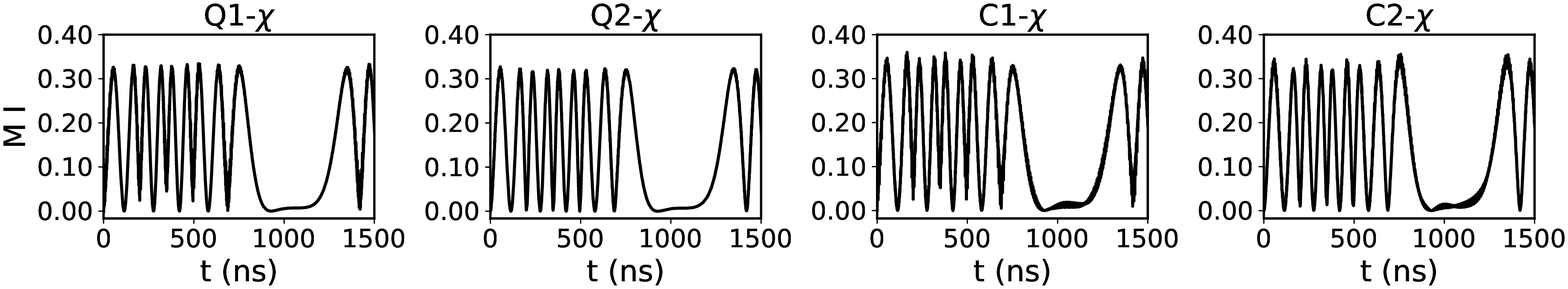}}
	\subfigure[]{\label{Fig:10b}\includegraphics[width=\textwidth]{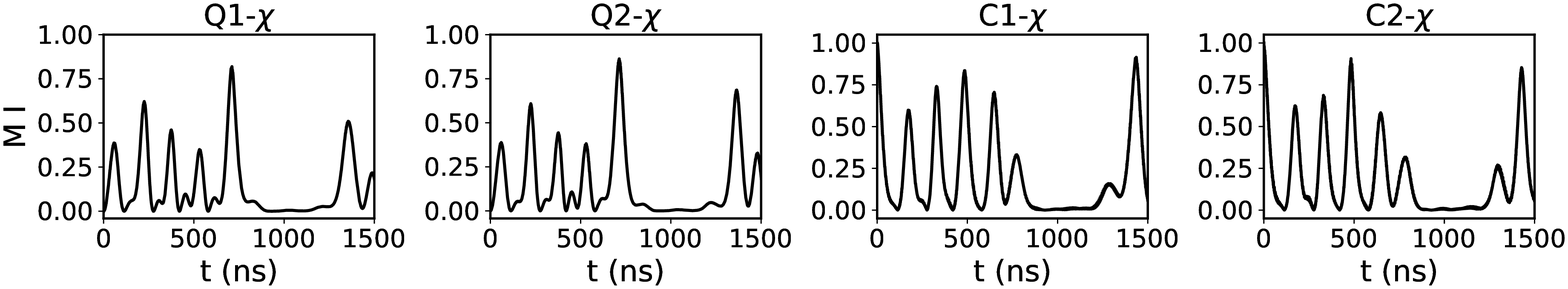}}
	\caption{\label{Fig:10}Mutual information for (a) $\ket{\psi(0)}=\ket{00100}$ and (b) $\ket{\psi(0)}=\frac{1}{\sqrt{2}}\left(\ket{00100}+\ket{01001}\right)$. }
\end{figure} 

\begin{figure}[h]
	\centering
	\subfigure[]{\label{Fig:11a}\includegraphics[width=\textwidth]{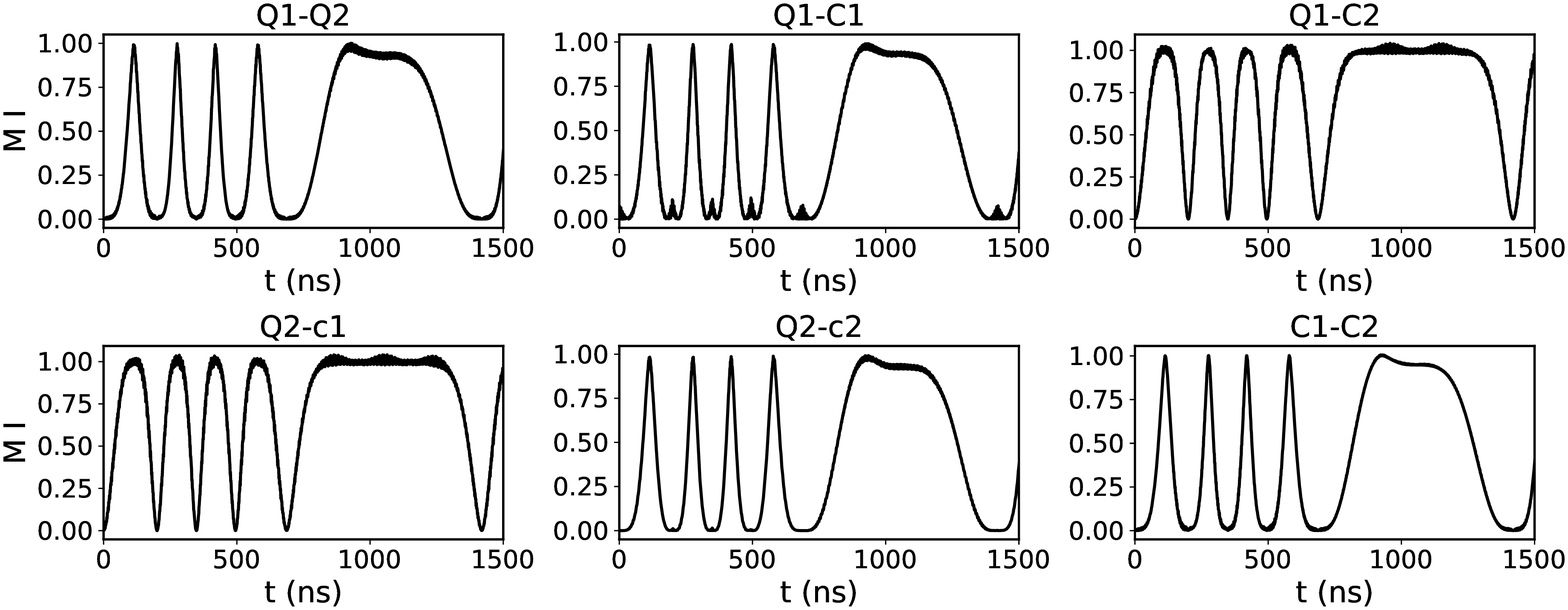}}
	\subfigure[]{\label{Fig:11b}\includegraphics[width=\textwidth]{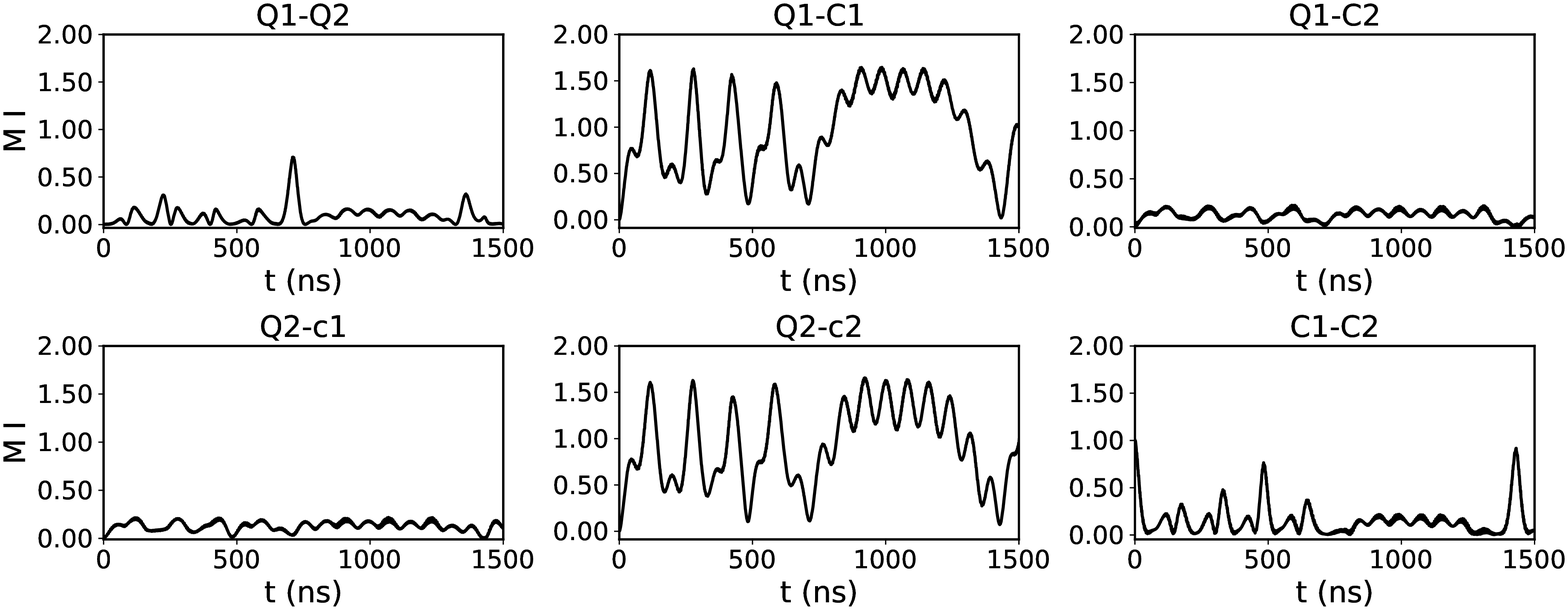}}
\caption{\label{Fig:11}Mutual information for (a) $\ket{\psi(0)}=\ket{00100}$ and (b) $\ket{\psi(0)}=\frac{1}{\sqrt{2}}\left(\ket{00100}+\ket{01001}\right)$. }
\end{figure}

Figures \ref{Fig:10a} and \ref{Fig:10b} shows the mutual information between the qubits and cavities with the $\chi^{(2)}$ bosonic mode for different initial conditions. As expected, the $\chi^{(2)}$ bosonic mode appears to share less information during the preservation time, since its coupling to the system has been reduced. We can see our initial entanglement between the cavity modes and the $\chi^{(2)}$ bosonic mode, in the second case as a non zero mutual information. Further, in Fig. \ref{Fig:11b}, we can also see an entanglement between the cavity 1 and 2. Thus, the entanglement we inferred from the von Neumann entropy is not bipartite. 

For cases 1 and 2, the qubits has a non zero mutual information between all other subsystems making a multipartite entanglement. And the time dependent coupling makes it possible to preserve it during the preservation slots. An ideal situation where the entanglement is completely preserved is when there is no interaction. Thus it is practically challenging to preserve it. Here, by using a time dependent non-linear coupling, we have preserved the entanglement produced in the configuration. 
 
\section{\label{sec:level6}Conclusion}

In this work, we have studied two coupled cavity with a single qubit in each and a second order nonlinear process in one of the cavity. The configuration for different values of nonlinear coupling and off resonant regimes are investigated and found that the system is highly sensitive to detuning. Further introduction of a harmonic time dependence on the nonlinear coupling has been utilized for the preservation of dynamics of the qubits. It is found that, a harmonic coupling with a control parameter, $\Omega$, can preserve the dynamics of the qubits for an interval irrespective of the choice of initial state. We further calculated the von Neumann entropy and mutual information to account for the entanglement present between the subsystem. From which it is found that, the preservation plateau also preserves the entanglement produced in the configuration.

\section*{Acknowledgement}

The authors, MTM and RBT, would like to thank the
financial support from KSCSTE, Government of Kerala
State, under Emeritus Scientist scheme. 
%%%%%%%%%%%%%%%%%%%%%%%%%%%%%%%%%%%%%%%%%%%%%%%%%%%%%%%%%%%%%%%%%%%%%%%%%%%%%%%%%%%

%\bibliographystyle{ws-jnopm}
\bibliographystyle{unsrt}
\bibliography{references}

\end{document}